\documentclass[aps,pra,twocolumn,showpacs,superscriptaddress]{revtex4}

\usepackage[latin1]{inputenc}
\usepackage[english]{babel}
\usepackage[dvips]{graphicx}
\usepackage{enumerate,amsthm,amsmath,amssymb,color,graphicx,bbm,float}
\usepackage{mathtools}
\usepackage{natbib}

\newtheorem{theo}{Theorem} 
\newtheorem{lemma}[theo]{Lemma}

\usepackage{epstopdf}

\usepackage{tikz}
\usetikzlibrary{arrows}

\pgfdeclareradialshading{myshading}{\pgfpointorigin}
{
  color(0mm)=(pgftransparent!0);
  color(2.5mm)=(pgftransparent!12);
  color(5mm)=(pgftransparent!39);
  color(7.5mm)=(pgftransparent!68);
  color(8.75mm)=(pgftransparent!78);
  color(10mm)=(pgftransparent!87);
  color(12.5mm)=(pgftransparent!96);
  color(20mm)=(pgftransparent!100)
}
\pgfdeclarefading{fading3}{\pgfuseshading{myshading}}

\newcommand{\be}{\begin{equation}}
\newcommand{\ee}{\end{equation}}

\newcommand{\scal}[2]{\left \langle #1 | #2 \right \rangle}
\newcommand{\scalmat}[3]{\left \langle #1 | #2 | #3 \right \rangle}

\def\multiset#1#2{\ensuremath{\left(\kern-.3em\left(\genfrac{}{}{0pt}{}{#1}{#2}\right)\kern-.3em\right)}}

\begin{document}

\title{Analysis of Imperfections in Practical Continuous-Variable Quantum Key Distribution}

\author{Paul Jouguet}
\affiliation{Institut Telecom / Telecom ParisTech, CNRS LTCI, 46, rue Barrault, 75634 Paris Cedex 13, France}
\affiliation{SeQureNet, 23 avenue d'Italie, 75013 Paris, France}

\author{S\'ebastien Kunz-Jacques}
\affiliation{SeQureNet, 23 avenue d'Italie, 75013 Paris, France}

\author{Eleni Diamanti}
\affiliation{Institut Telecom / Telecom ParisTech, CNRS LTCI, 46, rue Barrault, 75634 Paris Cedex 13, France}

\author{Anthony Leverrier}
\affiliation{Institute for Theoretical Physics, ETH Zurich, 8093 Zurich, Switzerland}

\date{\today}

\begin{abstract}
As quantum key distribution becomes a mature technology, it appears clearly that some assumptions made in the security proofs cannot be justified in practical implementations. This might open the door to possible side-channel attacks. We examine several discrepancies between theoretical models and experimental setups in the case of continuous-variable quantum key distribution. We study in particular the impact of an imperfect modulation on the security of Gaussian protocols and show that approximating the theoretical Gaussian modulation with a discrete one is sufficient in practice. We also address the issue of properly calibrating the detection setup, and in particular the value of the shot noise. Finally, we consider the influence of phase noise in the preparation stage of the protocol and argue that taking this noise into account can improve the secret key rate because this source of noise is not controlled by the eavesdropper.
\end{abstract}


\maketitle

Quantum Key Distribution (QKD) is a cryptographic primitive allowing two distant parties, Alice and Bob, to distill secret keys in an untrusted environment controlled by an eavesdropper, Eve \cite{SBC08}. Among quantum information technologies, QKD is one of the most advanced, and reaches already commercial applications. The main argument in favor of QKD is its provable security based on the laws of quantum mechanics; it is therefore particularly important to make sure that the security proofs derived for theoretical protocols can be applied to real-world implementations.
This is unfortunately never really the case because the security proofs usually assume idealized implementations, which do not take into account all possible experimental imperfections. This opens the door to potential security loopholes \cite{SK09} that might be successfully exploited by an attacker. Such side-channel attacks have already been demonstrated against commercial QKD systems \cite{XQL10,LWW10}.

There are basically two ways around side-channel attacks. A drastic solution consists in deciding that the systems held by Alice and Bob should not be trusted: this is the device-independent paradigm, based on the violation of a Bell inequality \cite{Eke91}. While being appealing in theory, this paradigm does not offer a practical solution since violating a Bell inequality in a loophole-free fashion has not been achieved until now.
A more practical way to address side-channel attacks aims at refining the theoretical models used for security proofs in order to include various sources of experimental imperfections. This involves, for instance, developing better models for the state preparation, including the light source, the modulation, and the noise, and for the detection, including the quantum efficiency and the calibration of the noise.

In this paper, we follow the second approach in the case of Continuous-Variable (CV) QKD protocols. The main specificity of these protocols is that they use a homodyne detection instead of single-photon counters, which makes them attractive from a practical perspective. Moreover, they are compatible with Wavelength Division Multiplexing \cite{QZQ10}, which is an important advantage when it comes to integrating QKD in real-world telecommunication networks.
CVQKD protocols are proven secure against coherent attacks \cite{RC09, FBB11, LGR12} and, asymptotically, the secret key rate is given by the Devetak-Winter formula \cite{DW05,BFS11} corresponding to collective attacks \cite{GC06,NGA06}. At the theoretical level, CVQKD protocols therefore present the same level of security as those based on photon counting, such as BB84 \cite{BB84}.

Here, we focus on Gaussian prepare-and-measure CVQKD protocols, which have already been demonstrated experimentally \cite{GVW03,LSS05,LBG07,QHL07,SAA07,FDD09,SSR10,MUL11} (see \cite{WPG12} for a recent review of all CVQKD protocols). In particular, we consider the GG02 protocol \cite{GG02} where Alice generates coherent states with a Gaussian modulation, and sends them to Bob who performs a homodyne measurement for a randomly chosen quadrature. By repeating this process a large number of times, Alice and Bob obtain correlated classical data, from which they can extract identical strings through the process of reconciliation \cite{JKL11} and then obtain a secret key using privacy amplification.

We study three kinds of imperfections that occur in all implementations of this protocol and see how they affect its security and the secret key rate. The first imperfection concerns the
modulation, which, in practice, can only approach the theoretical Gaussian modulation. Indeed, a Gaussian distribution is not only continuous but unbounded, and therefore cannot be exactly achieved since for instance, an infinite amount of randomness would be required. We show that the impact on security is not significant when the Gaussian distribution is replaced by a bounded, discrete approximation. However, deviations from a perfect discretized distribution degrade the security. The second source of imperfection comes from finite-size effects, and in particular from the calibration of the detection setup. While a first study in this direction has already considered statistical estimation of the transmittance and excess noise of the channel \cite{LGG10}, it assumed that the quantum efficiency and the electronic noise of the detection, and more importantly, the shot noise level, were all perfectly calibrated. Here, we consider these effects in detail and examine their impact on the secret key rate and distance.
Finally, we study the effect of phase noise in the preparation process of the protocol. This noise is unavoidable but one can safely assume that it is not controlled by the eavesdropper. We therefore show that by calibrating it properly, one can increase the secret key rate of the protocol.

The three kinds of imperfections are discussed in Sections I, II, and III of the paper, respectively.

\section{Security of Gaussian protocols with an imperfect modulation}

We first consider an issue present in all implementations of CVQKD with a Gaussian modulation, namely that it is impossible to use an exact Gaussian modulation in practice. In the ideal scenario for the prepare-and-measure protocol, for each signal to be sent, Alice is supposed to draw two random normal variables $q,p \sim \mathcal{N}(0,V_A)$ and to prepare the coherent state $|q + i p\rangle$ centered on the point $(q,p)$ in phase space.
Unfortunately, in practice, ignoring phase noise, the coherent state really prepared by Alice is centered on $(q',p') $ instead, where $(q',p')$ is a point on a finite grid, approximating the ideal value of $(q,p)$. This is unavoidable for several reasons. First, the analog-to-digital converters that drive the physical modulators used in practice produce discrete voltages; they typically have a bit depth of 10 like in \cite{FDD09}. Second, intensity modulators only work in some finite range of values, whereas the Gaussian distribution is unbounded. Another hardware constraint is the throughput of the physical Random Number Generators (for example Quantis, from ID Quantique, is limited to 16 Mbit/s). But there are also software limitations: one does not want to use too much randomness in order to draw the Gaussian variables $q$ and $p$ out of the uniform variables provided by the physical Random Number Generator because this requires computational power. For these reasons, it is useful to know how well the Gaussian modulation needs to be approximated in order to get a reasonably good level of security.

\begin{figure}
\begin{tikzpicture}
\draw[ultra thin,color=gray,step=.125cm]
(0,0) grid (3.5,3.5);

\draw[<->, ultra thin] (0.125,-0.1) -- (0.25,-0.1)
node[below] at (0.18725,0)%
{$\delta$};

\draw[<->] (0,-0.6) -- (3.5,-0.6)
node[below] at (1.75,-0.6)%
{$7\sqrt{V_A}$, $V_A=N_0$};

\draw[<->] (-0.6,0) -- (-0.6,0.5)
node[below] at (-1.15,0.55)%
{$\sqrt{N_0}$};

\draw[->] (-0.5,0) -- (4.0,0)
node[below right] {$q$};
\draw[->] (0,-0.5) -- (0,4.0)
node[left] {$p$};

\foreach \x in {0.5,1.0,1.5,2.0,2.5,3.0,3.5}
{
  \draw[ultra thin,dashed,color=gray] (-0.5,\x) -- (0,\x);
  \draw[ultra thin,dashed,color=gray] (\x,0) -- (\x,-0.5);
}

\pgfsetfading{fading3}{\pgftransformshift{\pgfpoint{1.5cm}{1.5cm}}}
\fill [black] (1.5,1.5) circle (4);
\end{tikzpicture}
\caption{Discretization grid used to approximate a Gaussian modulation in phase space. The modulation variance $V_A$ is chosen to be equal to the shot noise $N_0$. The distribution is truncated to $7$ standard deviations and discretized in steps of $1/4^{\text{th}}$ of shot noise units. A coherent state of variance $N_0$ covers a large part of the grid, which results in hiding the small imperfections of the discretized modulation.}
\label{figure:grid}
\end{figure}
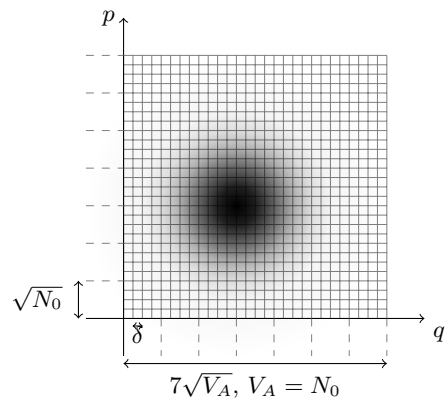

Intuitively, the presence of shot noise hides the small imperfections of the modulation and the security should not be compromised provided that the grid of $(q',p')$ is sufficiently fine-grained compared to the value of the shot noise. Figure \ref{figure:grid} illustrates how fine the grid needs to be compared to the shot noise.

In order to analyze the security of the practical protocol, it is convenient to look at the situation from Bob and Eve's points of view. In the theoretical protocol, the state sent by Alice to Bob should be a thermal state from Eve's perspective, that is a Gaussian mixture of coherent states. If Eve cannot distinguish the state sent in practice from a thermal state, then clearly the security of the protocol is not compromised by the approximated modulation. More precisely, if the trace distance between the ideal state and the actual state is bounded by $\epsilon_\mathrm{prep}$, and if the usual protocol (with perfect state preparation) is $\epsilon$-secure, then the true protocol is $(\epsilon+\epsilon_\mathrm{prep})$-secure. Therefore, one simply needs to ensure that $\epsilon_{\mathrm{prep}}$ can be made quite small, that is on the order of $10^{-10}$ in a realistic implementation.

\subsection{The quality of a Gaussian modulation}

Let us write $\displaystyle \rho = \rho_{\mathrm{th}} = \sum_{n=0}^\infty \frac{\overline{x}^n}{(\overline{x}+1)^{n+1}} |n\rangle\!\langle n |$ the ideal thermal state and $\displaystyle \sigma = \sum_k \omega_k |\alpha_k\rangle\!\langle \alpha_k|$ the state used in practice. Here $\omega_k$ corresponds to the probability of preparing the coherent state $|\alpha_k\rangle$.

We will compute the trace distance $|| \rho- \sigma||_1$ between the two states, for two discretizations $\sigma$, either with a cartesian or a polar grid. For both discretizations, we will use the gentle measurement lemma \cite{Win99,ON02}:
\begin{lemma}[Gentle measurement]
\label{gentle}
Let $\rho$ be a state and $\Pi$ be a projector. Then
\begin{equation}
||\rho - \Pi \rho \Pi|| \leq 2 \sqrt{1-\mathrm{tr} \left( \Pi \rho \Pi \right)}.
\end{equation}
\end{lemma}

Let us take $\Pi = |0\rangle\!\langle 0| + |1\rangle\!\langle 1| + \cdots + |Q-1\rangle\!\langle Q-1|$.
The triangle inequality gives:
\begin{eqnarray}
|| \rho - \sigma|| & \leq& || \rho - \Pi \rho \Pi || + || \Pi \rho \Pi - \Pi \sigma \Pi || + ||\Pi \sigma \Pi - \sigma||\nonumber\\
& \leq &\sum_{n=Q}^\infty \scalmat{n}{\rho}{n} +   \left| \sum_{n,m=0}^{Q-1} \scalmat{n}{\rho}{m} - \scalmat{n}{\sigma}{m} \right| \nonumber \\
&& + 2 \sqrt{1- \mathrm{tr} \left( \Pi \sigma \Pi \right) } \nonumber\\
& \leq &\sum_{n=Q}^\infty \frac{\overline{x}^n}{(\overline{x}+1)^{n+1}} + \sum_{n= 0}^{Q-1} \left| \frac{\overline{x}^n}{(\overline{x}+1)^{n+1}} - \scalmat{n}{\sigma}{n}\right| \nonumber\\
&&+2 \sum_{0 \leq n<m <Q} \left|\scalmat{n}{\sigma}{m} \right|  + 2 \sqrt{ 1- \sum_{n= 0}^{Q-1} \scalmat{n}{\sigma}{n}}\nonumber\\
&\leq &\left( \frac{\overline{x}}{\overline{x}+1}\right)^Q + \Delta_{\mathrm{diag}} + 2 \Delta_{\mathrm{nondiag}} + 2 \sqrt{R_{\sigma}}
\label{eq:initial_bound}
\label{eqsplit}
\end{eqnarray}
with $\displaystyle \Delta_{\mathrm{diag}} := \sum_{ 0\leq n <Q} \left| \frac{\overline{x}^n}{(\overline{x}+1)^{n+1}} - \scalmat{n}{\sigma}{n}\right|$,
$\displaystyle \Delta_{\mathrm{nondiag}} :=  \sum_{0 \leq n<m <Q} \left|\scalmat{n}{\sigma}{m} \right|$ and $\displaystyle R_{\sigma} := 1 - \sum_{0\leq n<Q} \scalmat{n}{\sigma}{n}$.
These three quantities can be estimated from the terms $\scalmat{n}{\sigma}{m}, 0 \leq n,m < Q$.

Notice that $R_{\rho} := \left( \frac{\overline{x}}{\overline{x}+1}\right)^Q$ does not depend on the actual approximation used, but only on the mean  photon number $\overline{x}$ of the ideal thermal state. When using a Gaussian modulation of variance $V_A$ (in shot noise units), one has $\overline{x}=2V_A$. This means that larger values of $V_A$ require larger values of $Q$ in order to obtain a good bound in Eq.~(\ref{eqsplit}). A typical range for $V_A$ is $[1, 20]$. For $V_A=20$, and $\epsilon_\mathrm{prep}=10^{-10}$, one needs to have $Q \approx 1000$ to ensure that $R_{\rho} \leq \epsilon_\mathrm{prep}$. Furthermore one also needs $R_{\sigma} \leq \epsilon_\mathrm{prep}^2$, which puts additional constraints on $Q$.

\subsection{Cartesian approximation}

Here, we consider an approximation of the form
\begin{align}
\sigma &= \sum_{k=-N}^{N} \sum_{l=-N}^{N} \omega_k \, \omega_l\, |\alpha_{kl}\rangle \: \langle \alpha_{kl} |
\end{align}
where $\omega_k = \frac{\gamma_k}{\sum_k \gamma_k}$, $\gamma_k= e^{-q_k^2/(2V)}$, $q_k = p_k = \frac{A}{N} k$, $\alpha_{kl}=q_k + i p_k$, and $A$, $N$ are two parameters to be optimized.
The $|\alpha_{kl}\rangle$ are coherent states: $|\alpha\rangle = e^{-|\alpha|^2/2} \sum_{n=0}^\infty \frac{\alpha^n}{\sqrt{n!}} \, |n\rangle$.

Therefore,
\begin{eqnarray*}
\scalmat{n}{\sigma}{m} &=& \sum_{k,l=-N}^{N} \omega_k \omega_l \scal{n}{\alpha_{kl}} \scal{\alpha_{kl}} {m} \\
&=& \sum_{k,l=-N}^{N} \omega_k \omega_l \, e^{-|\alpha_{kl}|^2} \frac{\alpha_{kl}^n \: {\alpha_{kl}^*}^m}{\sqrt{n! m!}}.
\end{eqnarray*}
From this expression, $\Delta_{\mathrm{diag}}$, $\Delta_{\mathrm{nondiag}}$ and $R_{\sigma}$ can be evaluated numerically for any choice of $\overline{x} = 2V_A, Q, A$ and $N$. Once $A$ is chosen, $N$ is typically set so that $\delta = A/N$, the discretization step, has some predetermined value. Given $V_A$, let us show that a low $\epsilon_\mathrm{prep}=|| \rho - \sigma||$ can be obtained with reasonable values of $A$ and $N$. Assume $V_A=20$, a rather large value corresponding to a Gaussian modulation of standard deviation $\sqrt{20}$; use (see Fig. \ref{figure:grid})
\begin{itemize}
\item
$A=7 \sqrt{V_A}$, meaning that the actual Gaussian distribution is truncated to 7 standard deviations;
\item
$N=\lceil 4A \rceil$, meaning that the distribution is discretized in steps of $1/4^{\text{th}}$ of shot noise units.
\end{itemize}
These choices can be used in practice: for $V_A=20$, they require $2 \times \lceil 4 \times 7 \times \sqrt{V_A} \rceil + 1 = 253$ discretization steps, that is, an 8-bit discretization grid. The entropy of the corresponding pair of discretized Gaussian values is $2 \times 6.2 = 12.4$ bits. Source coding techniques enable to use on average no more than this randomness quantity when drawing them in practice.

For $Q=2000$ (chosen to get a sufficiently low value of $R_{\sigma}$), a numerical evaluation yields
\begin{align}
\Delta_{\mathrm{diag}} &\leq 1.02 \ 10^{-11},\\
\Delta_{\mathrm{nondiag}} &\leq 1.04 \ 10^{-11},\\
R_{\sigma} &\leq 1.09 \ 10^{-24},
\end{align}
from which we deduce
\begin{equation}
||\rho -\sigma|| \leq 3.31 \ 10^{-11}.
\end{equation}

In the above discretization scheme, the mass lost because of the distribution cutoff is evenly distributed among the remaining coherent states. Let us give a similar result for a slightly different cutoff scheme where the lost mass is added to $\omega_{\pm N}$ only: $\omega_k = \frac{A}{N} \frac{1}{\sqrt{2 \pi V}} e^{-q_k^2/(2V)}$ for $-N+1 \leq i \leq N-1$, and $\omega_{-N}=\omega_N = (1 - \sum_{i=-N+1}^{N-1} \omega_k)/2$. For this scheme with the same parameters as before, we find
\begin{equation}
||\rho -\sigma|| \leq 2.98 \ 10^{-11}.
\end{equation}

\label{subs:cartesian}

\subsection{Polar approximation}

The actual modulation devices implement a polar modulation because phase and intensity are modulated separately. It is therefore natural to investigate the discretization required in polar coordinates to obtain a good approximation of a thermal state.

Let us assume that the polar coordinates are discretized uniformly on $[0,R] \times [0,2\pi]$.
Let us note the discretized values as:
\begin{align}
r_k &= \left( k +\frac{1}{2} \right) \frac{R}{K}, k \in [\![0,K-1]\!],\\
\theta_l &= \left( l +\frac{1}{2} \right) \frac{2\pi}{L}, l \in [\![0,L-1]\!].
\end{align}
We consider then an approximation of the form
\begin{align}
\sigma &= \frac{1}{L} \sum_k \omega_k \sum_l |\alpha_{kl}\rangle \: \langle \alpha_{kl} |,
\end{align}
where $\omega_k = \frac{\gamma_k}{\sum_k \gamma_k}$, $\gamma_k= r_k e^{-r_k^2/(2V)}$ and $|\alpha_{kl}\rangle = e^{-r_k^2/2} \sum_{n=0}^\infty \frac{r_k^n e^{in\theta_l}}{\sqrt{n!}} \, |n\rangle$.
Therefore,
\begin{align}
\scalmat{n}{\sigma}{m} &= \frac{1}{L} \sum_{k=0}^{K-1}\sum_{l=0}^{L-1} \omega_k \scal{n}{\alpha_{kl}} \scal{\alpha_{kl}} {m}\\
& = \frac{1}{L} \sum_{k=0}^{K-1}\sum_{l=0}^{L-1} \omega_k \, e^{-r_k^2} r_k^{n+m} \frac{e^{i(n-m)2\pi l/L}}{\sqrt{n! m!}}\\
& = U_{nm,L}\sum_{k=0}^{K-1}\omega_k \, \frac{e^{-r_k^2}r_k^{n+m}}{\sqrt{n! m!}}
\end{align}
with $U_{nm,L}=1$ if $L$ divides $n-m$, and $U_{nm,L}=0$ otherwise.

Unfortunately, this polar discretization requires a finer discretization than the cartesian one for the same approximation quality. For instance, with $V_A=20$ as before, using $Q=L=2000$ (thus eliminating the term $\Delta_{\mathrm{nondiag}}$ altogether in Eq.~(\ref{eq:initial_bound})) and $R=7 \sqrt{V_A}$, a 17-bit discretization of the amplitude is required to obtain $|\langle 0|\rho|0\rangle-\langle 0|\sigma|0\rangle| \leq 10^{-10}$. Drawing values corresponding to this discretization uses 11 bits for the angle and 15.5 bits for the modulus on average. This situation can be improved by using instead of regularly spaced $r_k$, points placed according to the Gauss quadratures method, especially the Gauss-Hermite variant: an 9-bit amplitude discretization entropy is found to be sufficient for $\epsilon_{\text{prep}} \leq 10^{-10}$. This is still slightly worse than the Cartesian grid, but could be improved further by making the angle discretization depend on the amplitude, as less points are needed in the vicinity of the origin.

\subsection{Robustness of bounds}

An important question related to the discretization is the robustness of the bounds given in the previous sections when the discretization grid is disturbed by some small systematic error term. This can happen, for instance, because of calibration errors or because of complex discretization effects due to the experimental setup. For example, an amplitude modulator generally produces an amplitude $A=\cos(cV+\phi)$, where $V$ is the voltage applied to it; since $V$ is discrete, the modulated amplitude values are projected to a set that is the image of the discrete set of attainable voltages by the functional realized by the modulator. To model the effect of these errors, we added a small disturbance with Gaussian distribution of standard deviation $\sigma_{\text{error}}$ to each point of the cartesian grid, and numerically computed the resulting $\epsilon_\mathrm{prep}$. With parameters as in Section \ref{subs:cartesian}, we get $||\rho-\sigma|| \approx 0.1 \times \sigma_{\text{error}}$. This shows that obtaining $\epsilon_\mathrm{prep} \leq 10^{-10}$ in practice may be difficult; it is more realistic to expect $\epsilon_\mathrm{prep} \approx 10^{-4}$ or $10^{-5}$.

It is true that the proof techniques used today force us to include $\epsilon_\mathrm{prep}$ in the final security parameter of the key, but it is plausible that this is too pessimistic. Indeed, it is known that protocols with a non-Gaussian modulation are secure against all attacks corresponding to a linear channel between Alice and Bob \cite{LG11}. This gives a hint that approximations of the order of $10^{-4}$ or $10^{-5}$ might be sufficient in practice.

\section{Imperfect Calibration of the Detection Setup}

We consider now finite-size effects related to the detection setup. We note that a proper calibration of Alice and Bob's devices is crucial to prove the security of the final key \cite{JWL11}. Our goal is to improve and expand the analysis of Ref. \cite{LGG10} concerning finite-size effects in CVQKD \footnote{Note that finite-size effects are also considered in Ref. \cite{FBB11}, where an entropic uncertainty relation is used to prove the security of an entanglement-based CVQKD protocol. Unfortunately, the bounds derived there are too pessimistic to be used in realistic experimental conditions.}.
In particular, the values of the quantum efficiency and the electronic noise of Bob's Homodyne Detection (HD) can only be estimated up to some finite precision. These inaccuracies must be taken into account when computing a secret key rate compatible with a realistic scenario (where these sources of noise are not assumed to be controlled by Eve) while considering finite-size effects. In the same way, the modulation variance on Alice's side and the excess noise on Bob's side both need to be estimated, in shot noise units, when computing the secret key rate. This implies that any imperfect precision on the estimation of the shot noise has an impact on the secret key rate.

The effect of a noisy HD has already been taken into account in the security proofs \cite{LBG07,FDD09}. The efficiency of the detection is modeled by a beamsplitter of transmittance $\eta$ and the electronic noise is modeled by a thermal noise of variance $N_\mathrm{el}$ added at the second input of the beamsplitter. That is, before Bob's HD, the state received by Bob is mixed with a thermal state of variance $N_\mathrm{el}$ on a beamsplitter of transmittance $\eta$. The variance of the electronic noise of the HD, $v_\mathrm{el}$, is linked to $N_\mathrm{el}$ by $v_\mathrm{el}=(1-\eta)(N_\mathrm{el}-1)$. Interestingly, the final key rate depends only on one parameter, namely the added noise referred to the input of the measurement device, denoted as  $\chi_\mathrm{hom}=\frac{1-\eta}{\eta}N_\mathrm{el}=\frac{1+v_\mathrm{el}}{\eta}-1$. Therefore, all the combinations of the parameters $(\eta, v_\mathrm{el})$ that give the same $\chi_\mathrm{hom}$ have the same impact on the secret key rate.

In \cite{FDD09}, these parameters were supposed to be calibrated in a secure lab, which implies that no attacker can interfere with the calibration procedure. Since this calibration is not performed during a QKD run, the statistical noise due to the finite number of samples used for the estimation can be made arbitrarily small. However, both parameters are still known imperfectly because of the finite precision of the measurement apparatuses. Here we consider an imperfect knowledge of these parameters and its effect on the secret key rate.

In order to calibrate a fiber-based HD, like the one used in \cite{FDD09}, one should in fact estimate three quantities:
\begin{itemize}
 \item the interferometer mode matching $\eta_\mathrm{mod}$ with precision $\Delta\eta_\mathrm{mod}$,
 \item the efficiency of the photodiodes $\eta_\mathrm{phot}$ with precision $\Delta\eta_\mathrm{phot}$,
 \item the fiber optic transmittance $\eta_\mathrm{opt}$ with precision $\Delta\eta_\mathrm{opt}$.
\end{itemize}
Then, the HD efficiency reads $\eta=\eta^2_\mathrm{mod}\eta_\mathrm{phot}\eta_\mathrm{opt}$ \footnote{Note that $\eta_\mathrm{mod}$ is derived from a measurement of the visibility of the interference fringes on one arm of the HD when the Local Oscillator (LO) interferes with another classical signal of the same intensity. It is therefore the experimentally useful quantity to characterize mode mismatching in the interferometer, and is used as a reference for modeling the equivalent beamsplitter transmittance.} and the overall uncertainty is:
\begin{equation}
\Delta\eta=\eta\left(2\frac{\Delta\eta_\mathrm{mod}}{\eta_\mathrm{mod}}+\frac{\Delta\eta_\mathrm{phot}}{\eta_\mathrm{phot}}+\frac{\Delta\eta_\mathrm{opt}}
{\eta_\mathrm{opt}}\right)
\end{equation}
The interferometer mode matching efficiency  $\eta_\mathrm{mod}$ is close to $99\%$, while a typical value for $\eta_\mathrm{phot}$ is $80\%$ with the PIN photodiodes used in \cite{FDD09}. The fiber optic transmittance is usually low (around $80\%$ for fiber-based HD since losses are usually applied on one arm of the interferometer to compensate for an unbalanced beamsplitter).

As far as $v_\mathrm{el}$ is concerned, this is estimated as the variance of the HD electronic noise, i.e., the detection output variance when no optical signal enters the detection device. This noise is mainly due to the thermal noise introduced by the load resistance at the entrance of the amplifier circuit (the intrinsic noise of the photodiodes is typically negligible). A straightforward way to determine $v_\mathrm{el}$ is to measure it directly as the variance of the HD output when no light enters the homodyne detection. Alternatively, one can plot the relationship between the power of a light source entering one branch of the beamsplitter of a balanced shot-noise limited HD and the variance of the HD output, when the other entrance of the HD is disconnected. This relationship should be linear, the Y-intercept being the variance of the electronic noise. Experimentally, the latter method leads to less accurate values of the electronic noise. However, even with the direct method $v_\mathrm{el}$ can only be known up to a precision $\Delta v_\mathrm{el}$.

The different uncertainties mentioned above can be evaluated depending on the measurement procedure and the precision of the measurement devices. In a practical CVQKD setup, Alice and Bob estimate the quantities required to compute the secret key rate through the sampling of $m=N-n$ pairs of correlated variables $(x_i,y_i)_{i=1\dots m}$, where $N$ is the total number of quantum signals sent through the quantum channel and $n$ is the number of signals used for the key establishment.

More precisely, the parameter estimation is performed in two steps.
First, after the state distribution and measurements, Alice and Bob need to roughly estimate the signal-to-noise ratio of their classical data in order to choose the proper error correcting code for the reconciliation \cite{JKL11}. This typically requires $m = O(\sqrt{N})$. Then, after the (reverse) reconciliation, Alice knows both her raw string and the one received by Bob. In practice, Alice and Bob would publicly compare a small hash of their final string to make sure that the reconciliation procedure succeeded. The size of these strings is $N$ and the parameter estimation can be performed \emph{on the whole string}. The results of this estimation will be used to compute a tight bound on Eve's information about Bob's string.

 Since for CVQKD, it is sufficient to estimate the covariance matrix of the state shared by Alice and Bob, the only parameters that need to be estimated are the variance on Alice's and Bob's sides, respectively $\langle x^2\rangle$ and $\langle y^2\rangle$, and the covariance between Alice and Bob $\langle xy\rangle$ (assuming here that $x$ and $y$ are centered variables, that is, that $\langle x\rangle = \langle y \rangle =0$). These values are linked to the key rate parameters through:
\begin{align}
 \langle x^2\rangle &= V_A\\
 \langle y^2\rangle &= \eta T V_A + N_0 + \eta T \xi + v_\mathrm{el}\\
 \langle xy\rangle &= \sqrt{\eta T}V_A,
\end{align}
where $T$ is the quantum channel transmittance, $V_A$ is the modulation variance, $\xi$ is the excess noise, and $N_0$ is the shot noise (all expressed in their respective units and not in shot noise units as it is usually assumed).

Since $\eta$ and $v_\mathrm{el}$ are calibrated beforehand, one has four unknown parameters $(V_A,N_0,T,\xi)$ and only three equations. However, by forcing a quantum
channel with zero transmittance, we get one more equation:
\begin{align}
 \langle {y_0}^2\rangle = N_0 + v_\mathrm{el}.
\end{align}
This can be done in Bob's laboratory by measuring the vacuum.

In order to compute confidence intervals for these parameters, we consider here a normal model for Alice and Bob's correlated variables $(x_i,y_i)_{i=1\dots N}$:
\begin{align}
 y=tx+z,
\end{align}
where $t=\sqrt{\eta T}\in \mathbb{R}$ and where $z$ follows a centered normal distribution with unknown variance $\sigma^2=N_0+\eta T\xi+v_\mathrm{el}$. Note that this normal model is an assumption justified in practice but not by \emph{current} proof techniques, which show that the Gaussian assumption is valid once the covariance matrix is known \cite{NGA06,GC06}. Exploiting symmetries of the protocol in phase-space might be a way to rigorously justify this assumption \cite{LKG09,LG10}. The random variable $x$ is a normal random variable with variance $V_A$ in the case of a Gaussian modulation. Another set of Bob's data $({y_0}_i)_{i=1\dots N'}$ can be used to measure the noise when no signal is exchanged (one can take $N'$ to be on the order of $N$):
\begin{align}
 y_0=z_0
\end{align}
where $z_0$ follows a centered normal distribution with unknown variance $\sigma_0^2=N_0+v_\mathrm{el}$.
Similarly to the analysis in \cite{LGG10}, Maximum-Likelihood estimators $\hat{t}$, $\hat{\sigma}^2$ and $\hat{\sigma_0}^2$ are known for the normal linear model:
\begin{align}
 \hat{t}&=\frac{\sum_{i=1}^N x_i y_i}{\sum_{i=1}^N x_i^2},\\
 \hat{\sigma}^2&=\frac{1}{N}\sum_{i=1}^N(y_i-\hat{t}x_i)^2,\\
 \hat{\sigma_0}^2&=\frac{1}{N'}\sum_{i=1}^{N'} {y_0}_i^2,\\
 \hat{V_A}&=\frac{1}{N}\sum_{i=1}^N {x}_i^2.
\end{align}
The estimators $\hat{t}$, $\hat{\sigma}^2$, $\hat{\sigma_0}^2$ and $\hat{V_A}$ are independent estimators whose distributions are:
\begin{align}
 &\hat{t} \sim \mathcal{N}\left(t,\frac{\sigma^2}{\sum_{i=1}^N x_i^2}\right),\\
 &\frac{N\hat{\sigma}^2}{\sigma^2}, \frac{N'\hat{\sigma_0}^2}{\sigma_0^2}, \frac{N\hat{V_A}}{V_A} \sim \chi^2(m-1)
\end{align}
where $t$, $\sigma^2$, $\sigma_0^2$ and $V_A$ are the true values of the parameters. In the limit of large $N,N'$, one can compute confidence intervals for these parameters:
\begin{align}
 t  & \in [\hat{t}-\Delta T,\hat{t}+\Delta T]\\
 \sigma^2 & \in [\hat{\sigma}^2-\Delta\sigma^2,\hat{\sigma}^2+\Delta\sigma^2]\\
 \sigma_0^2 & \in [\hat{\sigma_0}^2-\Delta\sigma_0^2,\hat{\sigma_0}^2+\Delta\sigma_0^2]\\
 V_A & \in [\hat{V_A}-\Delta V_A,\hat{V_A}+\Delta V_A],
\end{align}
where $\Delta T=z_{\epsilon_{PE}/2}\sqrt{\frac{\hat{\sigma}^2}{N V_A}}$, $\Delta\sigma^2=z_{\epsilon_{PE}/2}\frac{\hat{\sigma}^2\sqrt{2}}{\sqrt{N}}$,
$\Delta\sigma_0^2=z_{\epsilon_{PE}/2}\frac{\hat{\sigma_0}^2\sqrt{2}}{\sqrt{N'}}$, $\Delta V_A=z_{\epsilon_{PE}/2}\frac{\hat{V_A}\sqrt{2}}{\sqrt{N}}$ and
$z_{\epsilon_{PE}/2}$ is such that $1-\mathrm{erf}(z_{\epsilon_{PE}/2}/\sqrt{2})/2=\epsilon_{PE}/2$. Here we have used the error function $\mathrm{erf}(x)$, defined as:
\begin{align}
 \mathrm{erf}(x)=\frac{2}{\sqrt{\pi}}\int_0^x{e^{-t^2}dt}.
\end{align}
One can then estimate $T=\frac{\hat{t}^2}{\eta}$ and $\xi=\frac{\sigma^2-\sigma_0^2}{\hat{t}^2}$ using the previous estimators and their confidence intervals. As regards the shot noise, it is known with a precision that depends both on the number of samples used to compute the estimator $\hat{\sigma_0^2}$ and on the precision on the electronic noise $\Delta v_\mathrm{el}$.

Once the parameters and their respective confidence intervals have been determined, one can in particular express in shot noise units all the quantities needed to compute $S_{\epsilon_\mathrm{PE}}(y:E)$, the maximal value of the Holevo information between Eve and Bob's classical data compatible with the statistics except with probability $\epsilon_\mathrm{PE}$. Thus, the secret key rate for collective attacks including all the finite-size effects and calibration imperfections discussed previously can be computed as:
\begin{align}
K_{\text{finite}}=\frac{n}{N}(\beta I(x:y)-S_{\epsilon_\mathrm{PE}}(y:E)-\Delta(n)),
\end{align}
where $\beta I(x:y)$ is the amount of mutual information Alice and Bob were effectively capable to extract through the reconciliation phase ($\beta$ is the reconciliation efficiency which ranges from $0$ when no information was extracted to $1$ for a perfect reconciliation scheme) and $\Delta(n)$ is related to the security of the privacy amplification \cite{SR08,LGG10}.

\begin{figure}
\centering
 \includegraphics[width=80mm]{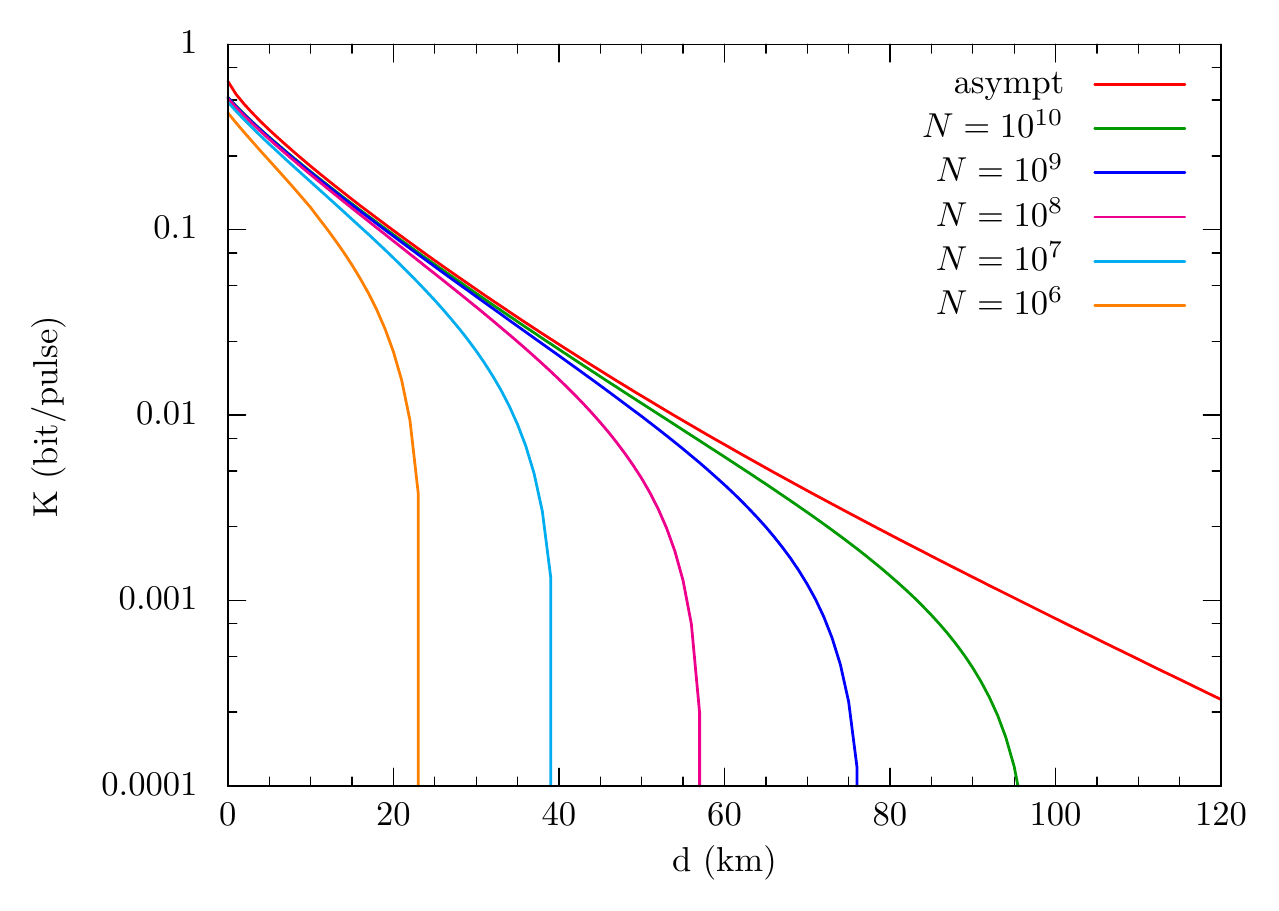}
  \caption{Secret key rate for collective attacks including finite-size effects and calibration imperfections with respect to the distance for different values of the number of samples. The transmittance $T$ and distance $d$ are linked with the expression $T = 10^{-\alpha d/10}$, where $\alpha$ is the loss coefficient of the optical fiber. $V_A=2.5$, $\xi=0.01$, $\eta=0.6$, $v_\mathrm{el}=0.01$, $\alpha=0.2$ dB/km, $\beta=95\%$, $\Delta\eta=0.1\eta$, $\Delta v_\mathrm{el}=0.1 v_\mathrm{el}$, $\epsilon=10^{-10}$, $N=\text{asympt},10^{10},10^9,10^8,10^7,10^6$ from top to bottom.}
   \label{figure:keyrate_finitesize}
\end{figure}

Figure \ref{figure:keyrate_finitesize} gives the secret key rate for various values of the number of samples $N=N'$. It appears that even taking pessimistic confidence intervals for $\eta$ and $v_\mathrm{el}$, for example with $\Delta\eta=0.1\eta$ and $\Delta v_\mathrm{el}=0.1 v_\mathrm{el}$, the impact on the secret key rate is not significant. However, a high precision on the shot noise is required for long distances since $\eta T \xi$ must be known with a high precision as already observed in \cite{LGG10}. It is worth noting that even using $10^6$ samples leads to a positive secret key rate for the Gaussian protocol unlike discrete modulation protocols for which at least $10^8$ samples are required \cite{LGG10}.

\section{Improved Key Rate with Phase Noise Calibration}

In order to obtain precise statements about the security of a given quantum key distribution (QKD) protocol, it is useful to carefully characterize the equipment of Alice and Bob. For CVQKD, this issue has already been addressed extensively for the detection stage. In particular, as was discussed in the previous section, in a calibrated device scenario, the detection model includes a finite quantum efficiency and a given level of electronic noise. Interestingly, both these imperfections act as sources of noise that can be trusted, in the sense that they are not controlled by Eve. This corresponds to the so-called realistic model, as opposed to the paranoid model where the eavesdropper is supposed to control all sources of noise. The realistic model allows one to derive a secret key rate that is actually better than the one obtained without this modeling for the imperfections of Bob's detection.

Concerning the preparation phase of the Gaussian CVQKD protocol that we are considering, recent work has addressed the issue of imperfections in Alice's state preparation. In particular, Refs \cite{Fil10,UF10,SPY11} studied the situation where Alice in fact prepares thermal states instead of coherent states. In fact, it is even possible to achieve CVQKD in the microwave regime where the preparation of pure coherent states is impossible \cite{WPL10,WPR11}. One remark about these works is that they consider a specific kind of imperfection that can be efficiently dealt with experimentally (at least in the optical regime). Indeed, if Alice really prepares thermal states instead of coherent states, one simple solution is to increase the variance of modulation and then to strongly attenuate the resulting state in order to obtain something very close to a coherent state. For this reason, the problem of preparing thermal states instead of coherent states is not really an issue in a practical scenario.

A more relevant issue concerns non-Gaussian sources of noise. In particular, there always is some phase noise on the state prepared by Alice. A typical value for the variance of this noise is $10^{-4} N_0$ per photon in the pulse \cite{LDT05}. One cannot suppress this noise by increasing the variance of the modulation and then attenuating the state, as mentioned above. Studying this noise is therefore of particular theoretical interest and of importance for actual experiments.

An important property of this noise is that it leaves the global state $\rho_{B_0} = \mathrm{tr}_A{\rho_{AB_0}}$ sent by Alice in the quantum channel (and therefore seen by Eve) invariant. This is different from the thermal noise considered in \cite{UF10,SPY11}, which increases the variance of $\rho_{B_0}$. In particular, this means that this noise can be modeled as an imperfect measurement for Alice in the entanglement-based equivalent protocol. In that picture, Alice prepares two-mode squeezed vacuum states, sends one mode to Bob and measures the other one with a heterodyne detection. When modeling the noise, one can keep the preparation of two-mode squeezed vacuum states, and only Alice's detection will be noisy. This simply means that the \emph{classical} data that she gets is noisy (with some phase noise). Therefore, the only consequence of this noise is that it degrades the mutual information shared between Alice and Bob, but it cannot increase Eve's information about Bob's measurement outcome, which is of interest in a reverse reconciliation scheme.

More specifically, in this case, the secret key rate against collective attacks is $K_{\text{asympt}} = \beta I(x:y) -\chi(y:E)$, where $\beta I(x:y)$ is defined as in the previous section and $\chi(y:E)$ is an upper bound on Eve's information on Bob's measurement outcomes. Because one can model phase noise as a local noise acting on Alice's system, it can only decrease the quantity $I(x:y)$ but cannot help the eavesdropper by increasing $\chi(y:E)$. In such a scenario, one can expect that the phase noise can be removed from the excess noise when computing Eve's information, leading to a realistic model for the preparation stage, similarly to the detection stage. This should lead to better secret key rates in practice.

\subsection{Model for the phase noise}

The phase noise can be modeled as applying a phase rotation $U(\theta)= \exp (i \theta a^\dagger a)$ on Alice's mode with a random phase $\theta$ characterized
by some probability distribution $p(\theta)$.
This means that when Alice tries to prepare some coherent state $|\alpha\rangle$ in the prepare-and-measure protocol, she actually prepares a state with a noisy phase: $\rho_\alpha = \int U(\theta) |\alpha\rangle\!\langle \alpha|U(\theta)^\dagger p(\theta) \mathrm{d} \theta$.
Let us assume that Alice initially prepares an ideal two-mode squeezed vacuum state with a variance $V_A$. This state $\rho_\mathrm{ideal}$ has the following
covariance matrix (for a displacement vector $[q_A, p_A, q_B, p_B]^T$):
\begin{equation}
\Gamma_\mathrm{ideal} =
\left[
\begin{array}{cc}
V_A \mathbbm{1}_2 & W \sigma_z  \\
W \sigma_z &  V_A \mathbbm{1}_2\\
\end{array}
\right],
\end{equation}
where $W := \sqrt{V^2_A-1}$ and $\sigma_z = \mathrm{diag}(1,-1)$.

Applying a local phase shift $U(\theta)$ on Alice's mode gives a state with a covariance matrix $\Gamma(\theta)$ given by
\begin{equation}
\Gamma(\theta) = \left[
\begin{array}{cccc}
V_A & & W \cos \theta & W \sin \theta\\
& V_A & W \sin \theta & -W \cos \theta \\
W \cos \theta & W \sin \theta& V_A & \\
W \sin \theta & -W \cos \theta & &V_A \\
\end{array}
\right].
\end{equation}

Finally, the state affected by the phase noise is a classical mixture of states with random phase shifts
$\rho = \int (U_A(\theta) \otimes \mathbbm{1}_B) \rho_\mathrm{ideal} (U_A(\theta)^\dagger \otimes \mathbbm{1}_B) p(\theta) \mathrm{d} \theta$, and its covariance matrix is
\begin{equation}
\Gamma_\mathrm{phase\, noise} =
\left[
\begin{array}{cc}
V_A \mathbbm{1}_2 & \sqrt{\kappa} W  \sigma_z  \\
\sqrt{\kappa}W  \sigma_z &  V_A \mathbbm{1}_2\\
\end{array}
\right],
\end{equation}
where we assumed that the distribution $\theta$ is symmetric, and more precisely that $\int p(\theta) \sin \theta \mathrm{d}\theta =0$, and introduced $\kappa := \left( \int p(\theta) \cos \theta\mathrm{d} \theta\right)^2 = (E[\cos \theta])^2$, where $E[X]$ is the expectation of the random variable $X$.

The interesting point is that from both Bob and Eve's points of view, it does not change anything whether a random phase shift is applied. In particular, the value of $\chi(y:E)$ quantifying the information that Eve can acquire about the raw key in a \emph{reverse reconciliation} scenario does not depend on the value of the phase noise. Note that this statement would not be true in a direct reconciliation scenario where the raw key would correspond to Alice's noisy data.

Let us suppose that the quantum channel between Alice and Bob is characterized by its transmittance $T$ and excess noise $\xi$.
The covariance matrix $\Gamma_{AB}$ of the bipartite state shared by Alice and Bob after the quantum channel is then given by:
\begin{equation}
\Gamma_{AB}=
 \left[
\begin{array}{cc}
V_A \mathbbm{1}_2 & \sqrt{\kappa T} W \sigma_z\\
\sqrt{\kappa T} W \sigma_z  & (T(V_A-1) + 1 + T \xi)\mathbbm{1}_2 \\
\end{array}
\right].
\end{equation}

If they were not taking phase noise into account (that is, if $\kappa$ was equal to 1), Alice and Bob would estimate a transmittance $T'$ and an excess noise $\xi'$ such that
\begin{equation}
\left\{\begin{array}{ll}
&T'  = T \kappa\\
&T'(V_A-1) + 1 + T' \xi' = T (V_A-1) + 1 + T \xi
\end{array}\right.
\end{equation}
that is
\begin{equation}
\left\{\begin{array}{ll}
T &= T'/ \kappa \\
\xi & =  \xi' - (1-\kappa)(V_A-1)
\end{array}\right.
\end{equation}

If the phase noise parameter $\kappa$ is known, one can estimate the covariance matrix as usual, hence obtaining values $(T', \xi')$ and use the formula above to deduce the parameters $(T, \xi)$ that can be used instead to compute Eve's information $\chi(y:E)$.

For this technique to work, it is necessary to be able to measure $\kappa=(E[\cos \theta])^2$ experimentally. This is discussed in the next section.

\subsection{Experimental evaluation of the phase noise}

The evaluation of the phase noise can be performed with a phase sensitive apparatus which allows us to compute an estimate of the noise between a signal whose quadratures are modulated following a chosen sequence and the outputs of some chosen quadrature measurements. A homodyne or heterodyne detection can be used for this purpose.

Similarly to what is done on Bob's side when the homodyne detection efficiency and the variance of the electronic noise are calibrated, it is necessary to assume that the calibration of the phase noise is performed in a safe place, i.e. that Eve cannot interfere with Alice's apparatus during the phase noise measurement.
The measurement can also be performed during a run of the protocol but one still needs to assume that Eve cannot interfere with Alice's device. This is crucial since overestimating the phase noise would lead to an overestimation of the secret key rate.

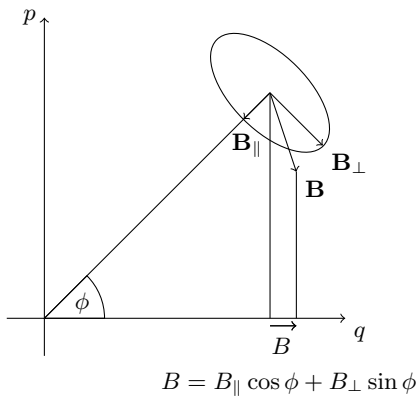
\begin{figure}
\begin{tikzpicture}
\draw[->] (-0.5,0) -- (4.0,0)
node[below right] {$q$};
\draw[->] (0,-0.5) -- (0,4.0)
node[left] {$p$};

\draw[thin]
(0,0) -- (8mm,0mm) arc (0:45:8mm) -- cycle
node at (5mm,2mm) {$\phi$};

\draw (0,0) -- (3,3);

\draw[rotate around={135:(3,3)},black] (3,3) ellipse [x radius=1 cm, y radius=0.5 cm];

\draw[->] (3,3) -- (3-0.35,3-0.35)
node at (2.7,2.3) {$\mathbf{B_{\parallel}}$};

\draw[->] (3,3) -- (3+0.7,3-0.7)
node[below right] {$\mathbf{B_{\perp}}$};

\draw[->] (3,3) -- (3+0.7-0.35,3-0.7-0.35)
node[below right] {$\mathbf{B}$};

\draw (3,3) -- (3,0);
\draw (3+0.7-0.35,3-0.7-0.35) -- (3+0.7-0.35,0);

\draw[->] (3,-0.1) -- (3+0.7-0.35,-0.1)
node[below] at (3.15,-0.15) {$B$};
\draw[->] (3,-0.1) -- (3+0.7-0.35,-0.1)
node[below] at (3.25,-0.6)
{$B=B_{\parallel}\cos{\phi}+B_{\perp}\sin{\phi}$};
\end{tikzpicture}
\caption{Experimental evaluation of the phase noise. The noise $\mathbf{B}$ before Bob's measurement can be decomposed into the sum of a component orthogonal
to the signal $\mathbf{B_{\perp}}$ and a component parallel to the signal $\mathbf{B_{\parallel}}$. The result of Bob's $q$-quadrature measurement is $A\cos{\phi}+B$ where Alice prepared the state centered in $Ae^{i\phi}$ with $A\geq 0$.}
\label{figure:phase_noise}
\end{figure}

Here, we are interested in the phase noise in the prepare-and-measure version of the protocol.
The procedure to estimate it goes as follows: Alice modulates as usual with a bivariate Gaussian distribution and she measures either one of the quadratures
with a homodyne detection. Computing the variance of her measurement outcomes allows here to infer the quantity $\kappa$ introduced above.
Let us denote by $\phi$ the random variable corresponding to the angle between the modulated state and the measured quadrature and by $B$ the random variable
corresponding to the noise. This means for example that Alice prepared the state centered in $A e^{i\phi}$ (with $A \geq 0$) and that the outcome of her
$q$-quadrature measurement was $A \cos \phi + B$. This noise $B$ can be decomposed into the sum of a component orthogonal to the signal and a component parallel to the signal:
\begin{align}
 B=B_{\parallel}\cos{\phi}+B_{\perp}\sin{\phi},
\end{align}
where we assume that $B_{\parallel}$ and $B_{\perp}$ are independent of $\phi$. Figure \ref{figure:phase_noise} gives an illustration of this decomposition.
We can easily build estimators of $B_{\parallel}$ and $B_{\perp}$ (in the following, $E[X]$ and $V[X]$ refer respectively to the expectation and variance of the random variable $X$):
\begin{align}
V[B \cos{\phi}] &= V[B_{\parallel} \cos^2{\phi}] + V[B_{\perp} \cos{\phi} \sin{\phi}]\\
&= V[B_{\parallel}] E[\cos^4{\phi}] + V[B_{\perp}] E[\cos^2{\phi} \sin^2{\phi}]\\
&= 3/8 V[B_{\parallel}] + 1/8 V[B_{\perp}]\\
V[B \sin{\phi}] &= 1/8 V[B_{\perp}] + 3/8 V[B_{\parallel}]
\end{align}
Since both $V[B \cos{\phi}]$ and $V[B \sin{\phi}] $ can be measured experimentally, one therefore has access to the values of $V[B_{\perp}] $ and $V[B_{\parallel}]$.
Here, we assume that the only sources of noise are the shot noise and the phase noise. We assume that $B_{\perp}$ can be fully described by the shot noise and the phase noise:
\begin{align}
 V[B_{\perp}] &= N_0 + V[A\sin{\theta}]=N_0 + E[\sin^2{\theta}]E[A^2]\\
 E[\sin^2{\theta}] &= \frac{V[B_{\perp}] - N_0}{E[A^2]}=E_1
\end{align}
where $A$ is the amplitude of the modulated signal and where we used $E[\sin{\theta}]=0$.
The assumption of a small phase noise, i.e. small values of $\theta$, gives:
\begin{align}
 E[\cos{\theta}] &= E[1-\theta^2/2]\\
&= 1 - \frac{1}{2}E_1
\end{align}

\begin{figure}
\centering
 \includegraphics[width=80mm]{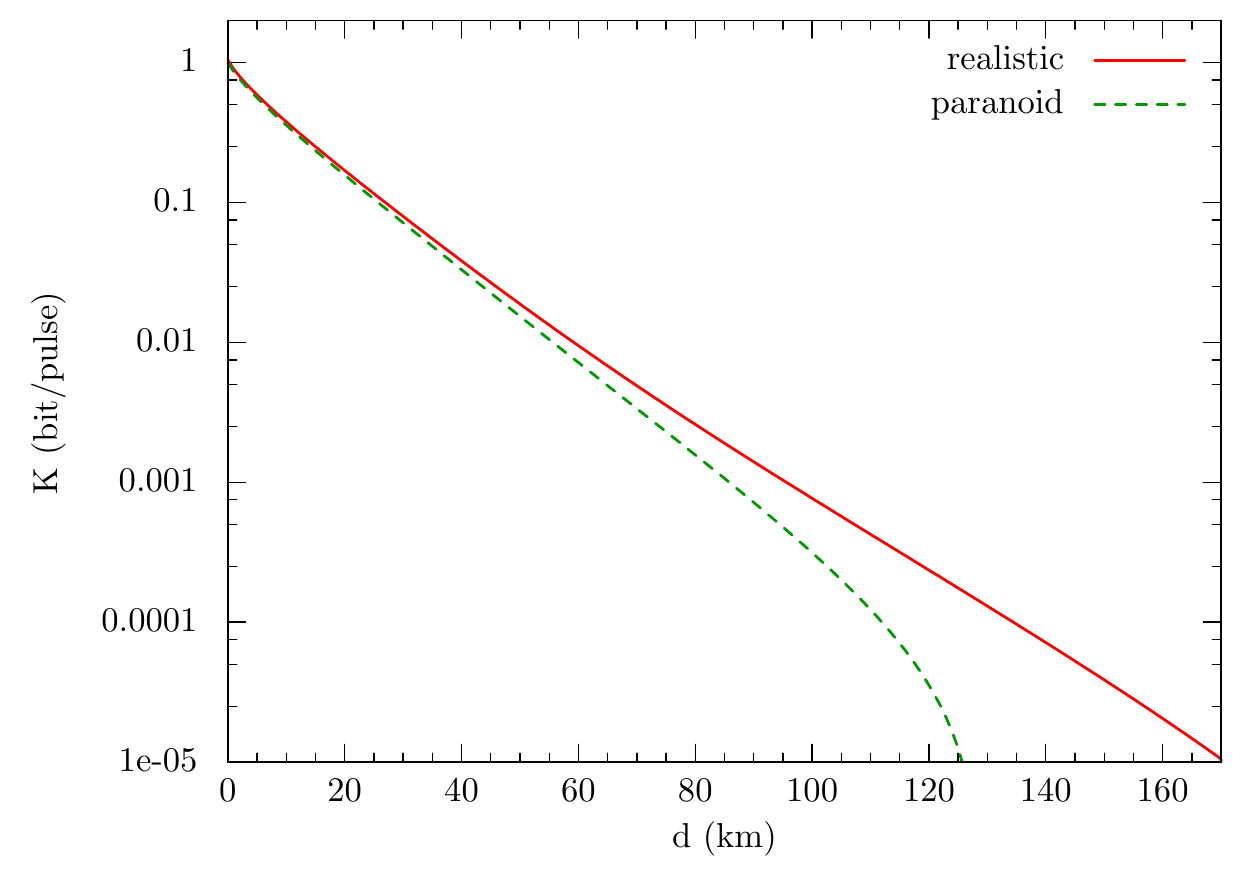}
  \caption{Secret key rate for collective attacks in the asymptotic regime. The plot at the top is obtained in the so-called realistic model where the phase noise is calibrated and is considered as a local noise useless to the eavesdropper. The plot at the bottom corresponds to the so-called paranoid model where all the sources of noise are attributed to the eavesdropper. The transmittance $T$ and distance $d$ are linked with the expression $T = 10^{-\alpha d/10}$, where $\alpha$ is the loss coefficient of the optical fiber. $V_A=2.5$, $\xi=0.025$, $\alpha=0.2$ dB/km, $\beta=95\%$, $E_1=3 \ 10^{-3}$.}
   \label{figure:keyrate_phasenoise}
\end{figure}

Figure \ref{figure:keyrate_phasenoise} compares the so-called realistic and paranoid models. We consider a pessimistic scenario where the excess noise on Alice's side is about $2.5\%$ of the shot noise (the detector quantum efficiency and electronic noise are not taken into account here, for clarity). For a modulation variance $V_A=2.5$, we measured experimentally $E_1=3 \ 10^{-3}$ with a system similar to the one described in \cite{FDD09}. This leads to a realistic value of the excess noise $\xi_{\text{real}}=1.75\%$. The result on the secret key rate for collective attacks is an increased achievable distance by about $40$ km.

\section{Conclusion}
In this work, we have analyzed several types of imperfections that appear in practical implementations of Gaussian continuous-variable QKD protocols. In particular, we studied a realistic approximate Gaussian modulation in the state preparation at Alice's site, the calibration of detection characteristics estimated with a finite precision at Bob's site, and the presence of intrinsic phase noise in the prepared states. In all cases, we provided a precise model of the imperfection and used this model to examine its effect on the security and performance of the protocol. These effects are more or less significant in practice: it is clear, for instance, that taking into account the phase noise in the security proof of a realistic scenario provides an important advantage in terms of secret key rate, while carefully approximating the ideal Gaussian modulation with respect to the shot noise values can minimize the impact of this imperfection. Finally, as expected, finite-size effects at all stages of the protocol should always be considered when calculating practical secret key rates.

This analysis demonstrates the importance of refining security proofs of QKD protocols to consider practical imperfections. In particular for CVQKD protocols, where potential side channels have not been yet widely studied, it provides specific ways to bypass attacks based on improperly modeled devices and procedures.

\section{Acknowledgements}
This research was supported by the French National Research Agency, through the FREQUENCY (ANR-09-BLAN-0410) and HIPERCOM (2011-CHRI-006) projects, and
by the European Union through the project Q-CERT (FP7-PEOPLE-2009-IAPP). P. Jouguet acknowledges support from the ANRT (Agence Nationale de la Recherche
et de la Technologie). A.L. was supported by the SNF through the National Centre of Competence in Research ``Quantum Science and Technology''.

\bibliography{imperfect}

\end{document}